\documentclass[prb,aps,twocolumn,showpacs,nobibnotes,epsf]{revtex4}
\usepackage{graphicx}% Include figure files
\usepackage{dcolumn}% Align table columns on decimal point
\usepackage{bm}% bold math
\usepackage{SIunits}

\begin{document}
\title{ Rearrangement of Sodium ordering and its effect on physical properties in $Na_xCoO_2$ system}
\author{T. Wu, K. Liu, H. Chen, G. Wu, Q. L. Luo, J. J. Ying}
\author{X. H. Chen}
\altaffiliation{Corresponding author} \email{chenxh@ustc.edu.cn}
\affiliation{Hefei National Laboratory for Physical Science a
Microscale and Department of Physics, University of Science and
Technology of China, Hefei, Anhui 230026, People's Republic of
China\\}

\begin{abstract}
We systematically study Raman spectroscopy of cleaved Na$_x$CoO$_2$
single crystals with 0.37 $\leq$ x $\leq$ 0.80. The Raman shift of
A$_{1g}$ mode is found to be linearly dependent on Na content, while
the Raman shift of E$_{1g}$ mode has an abnormal shift to high
frequency around x = 0.5. The abnormal shift is ascribed to the
occurrence of Na rearrangement in O1 structure. Temperature
dependent Raman spectrum for x = 0.56 sample shows that Na
rearrangement transition from O1 structure to H1 structure occurs
around 240 K. Electronic transport and susceptibility for the sample
with $x=0.56$ show a response to the Na rearrangement transition
from O1 to H1 structure, and that different Na ordering pattern
causes distinct physical properties. These results give a direct
evidence to proved Na ordering effect on physical properties of Co-O
plane.

\end{abstract}


\vskip 300 pt

\maketitle

\section{Introduction}

  The layered cobaltate Na$_x$CoO$_2$ has attracted much interesting in
strong correlation research area due to its rich
physics.\cite{Takada, Ong, Foo} As we know, the electronic ground
state is strongly dependent on the Na content in
Na$_x$CoO$_2$\cite{Foo} in which the valence of Co ion can be tuned
by Na content. Moreover, the change of Na content can also affect
the structure of Na$^+$ ion layer. Early electron diffraction and
neutron diffraction measurements reveal a kaleidoscope of Na$^+$ ion
patterns as a function of concentration.\cite{Zandbergen, Huang,
Huang2, Huang3} A detailed phase diagram of Na$^+$ ion layer
structure was given by Huang et al..\cite{Huang2} As shown in Fig.1,
there are four different types of structure for Na$^+$ ion layer
with different Na content --- H1, H2, H3 and O1. In the four types
of structure, there are three types of Na$^+$ ion site--- Na(1)(2b
(0, 0, 1/4)), Na(2)(2c (2/3, 1/3, 1/4)) and Na(2)'(6h (2x, x, 1/4)).
The occupation probability of the above sites depends on both Na
content and temperature.\cite{Huang2} It is expected that different
Na$^+$ ion pattern corresponds to distinct physical properties. The
relationship between structure of Na$^+$ ion layer and physical
properties of Co-O plane is believed to be an important key of
understanding this system\cite{Foo, Huang2} and becomes a hot issue
in recent researches\cite{Roger, Schulze, Morris, Chou, Balicas,
Pai, Zhou, Marianetti, Choy}.
\begin{figure}[t]
\centering
\includegraphics[width=0.4\textwidth]{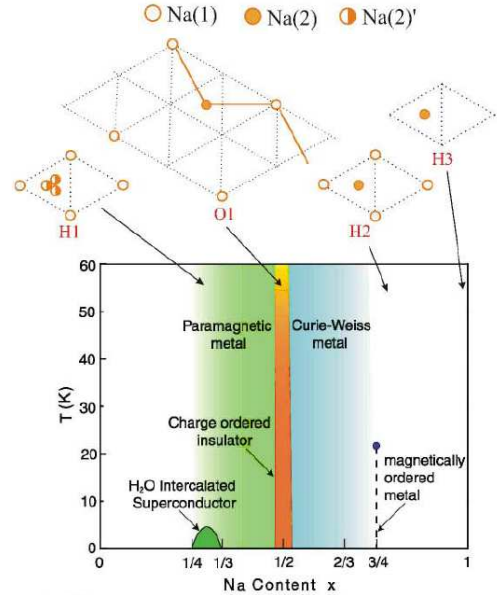}
\caption{(Color online). Correspondence between structure and
properties in Na$_x$CoO$_2$(Figure comes from ref[6]). Upper
panel: schematic of the Na ion distributions in the four
Na$_x$CoO$_2$ phases; Lower panel: the electronic phase diagram.}
\label{fig1}
\end{figure}

Recently, single crystal neutron scattering study reveals the
formation of ordered sodium vacancies for x$>$0.5. Inside each
vacancy, there exists a Na monomer or a Na trimer occupying Na(1)
sites that sit atop Co atoms.\cite{Roger} This multi-vacancy model
and Na trimer geometry are also proved by single crystal synchrotron
X-ray diffraction and can be used to explain the observed
superstructures of x=0.84 and x=0.71 crystals very well.\cite{Chou}
A recent STM study also found that an unexpected Na trimer ordering
is found for x$\leq$0.5 samples.\cite{Pai} These interesting
findings make us renew our understanding of Na ordering pattern.
More and more theoretical and experimental progress of Na
ordering\cite{Roger, Chou, Zhou, Marianetti, Choy} show that Na
ordering plays an important role in understanding of the rich
physical properties of Co-O plane. However, direct experimental
evidence of relationship between Na ordering and physical properties
of Co-O plane is still limited\cite{Schulze}. Here, we report a
systematically Raman study in Na$_x$CoO$_2$. It is found that an
abnormal sodium dependence of E$_{1g}$ mode occurs around x=0.5.
Temperature dependent Raman study indicates that a possible Na
rearrangement transition from H1 to O1 phase defined by Huang et
al.\cite{Huang2} occurs at x=0.56. Electronic transport and
susceptibility for the sample with x=0.56 show that this kind of Na
rearrangement transition makes different physical properties in this
system. These results give a direct evidence for Na ordering effect
on the physical properties of Co-O plane.

\section{Experiment}

  High quality single crystals of Na$_x$CoO$_2$ were grown using
the flux method (x = 0.7) and floating zone technique (x = 0.75 and
0.80). The Na$_x$CoO$_2$ sample with x$<$0.7 is prepared by sodium
deintercalation of the Na$_{0.7}$CoO$_2$ singe crystals in solutions
obtained by dissolving I$_2$ (0.2 M, 0.02 M, 0.004M) or Br$_2$ (6.0
M) in acetonitrile (M is "molar") for 4 days at room temperature.
The x values of the samples were estimated by the same method as
previous paper.\cite{Wu} Raman spectra were obtained on a LABRAM-HR
Confocal Laser MicroRaman Spectrometer using the 514.5 nm line from
an argon-ion laser with in-plane light polarization. The single
crystals were cleaved to obtain fresh surface before Raman
measurements. The resistance was measured by an AC resistance bridge
(LR-700, Linear Research). Magnetic susceptibility measurements were
performed with a superconducting quantum interference device
magnetometer in a magnetic field of 7 T. It should be addressed that
all results discussed as follow are well reproducible.

\section{Result and Discussion}

\begin{figure}[h]
%\captionstyle{flushleft} \onelinecaptionsfalse
\centering
\includegraphics[width=0.4 \textwidth]{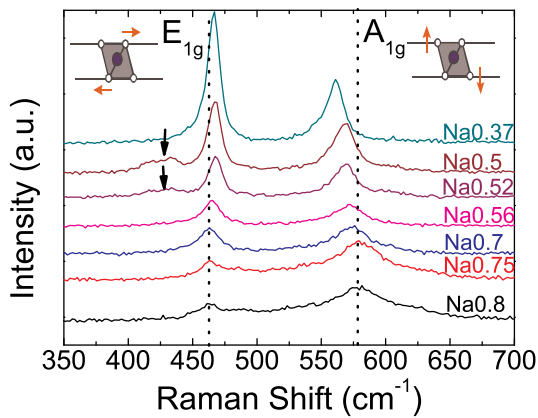}
\caption{(color online). Raman spectra of Na$_x$CoO$_2$ single
crystal from ab-plane with x=0.37, 0.50, 0.52, 0.56, 0.70, 0.75,
0.80 at room temperature.} \label{fig2}
\end{figure}

  Fig.2 shows the Raman spectra for x=0.37, 0.50, 0.52, 0.56, 0.7,
0.75 and 0.80 at ambient temperature with the range from 350 to
700 cm$^{-1}$. Raman spectra of all samples are measured from
ab-plane with in-plane light polarization. In Fig.2, all samples
show two distinct Raman modes around 467 cm$^{-1}$ and 561
cm$^{-1}$, which can be attributed to Raman-active in-plane
E$_{1g}$ and out-of-plane A$_{1g}$ vibrations of oxygen in CoO$_6$
octahedra, respectively.\cite{Shi, Iliev, Lemmens, Qu, Zhang,
Lemmens2} The samples with x= 0.5 and 0.52 show additional modes
around 425 cm$^{-1}$, which is attributed to Na$^+$ ion
ordering.\cite{Qu} In addition, both of two samples show a charge
ordering behavior at low temperature which is consistent with the
results by Foo et al.\cite{Foo} and Wu et al..\cite{Wu} These
results indicate the sensitivity of this phonon to structural and
electronic ordering processes. Two distinct systematic shifts are
observed for A$_{1g}$ and E$_{1g}$ modes. With decreasing Na
content, the A$_{1g}$ mode shifts to low frequency and E$_{1g}$
mode shifts to high frequency. The widths of A$_{1g}$ and E$_{1g}$
modes increase with sodium doping, and there exists a distinct
broadening of A$_{1g}$ and E$_{1g}$ modes above x$\sim$0.52. This
broadening of the two modes maybe come from instability of
ordering structure of Na$^{+}$ ions. The systematic changes of
related Raman shift of A$_{1g}$ and E$_{1g}$ with decreasing Na
content are shown in Fig.3. The Raman shift of A$_{1g}$ mode is
linearly suppressed with Na content. This result is consistent
with previous data.\cite{Lemmens2, Donkov} But it is surprising
that the Raman shift of E$_{1g}$ mode presents an abnormal shift
around x = 0.5, and this phenomenon is not observed in previous
data.\cite{Donkov} Huang et al. have proposed a structural phase
diagram based on different structure of Na$^+$ ion
layer.\cite{Huang2} Present data indicate that the O1 structure
may be corresponding to the abnormal shift of E$_{1g}$ mode.
Meanwhile, above results also indicate that the E$_{1g}$ mode is
more sensitive to O1 structure than A$_{1g}$ mode, which can be
used as the fingerprint of O1 structure. It should be emphasized
that the temperature dependence of Raman shift for E$_{1g}$ mode
also shows a T-linear behavior except for that observed in the
samples of $x \sim 0.5$ with Na ordering in O1 structure. Since
the in-plane Na$^+$ ion ordering happens between two neighboring
CoO$_6$ layers, it is also easy to understand why the out-of-plane
A$_{1g}$ mode has no distinct change. Recently, multi-vacancy
model was proposed to understand the Na ordering
phenomenon\cite{Roger} and Na rearrangement phenomena were also
found for x$>$0.5.\cite{Huang3, Morris} The above results suggest
that a possible Na rearrangement transition occurs around x=0.5
and E$_{1g}$ mode is a good indicator.

\begin{figure}[h]
%\captionstyle{flushleft} \onelinecaptionsfalse
\centering
\includegraphics[width=0.4\textwidth]{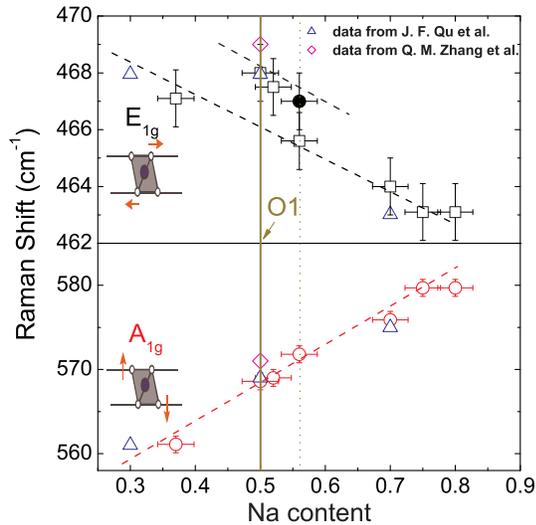}
\caption{(color online). Na content dependent Raman shift of
A$_{1g}$ and E$_{1g}$ mode at ambient temperature. Open squares
stand for E$_{1g}$ mode. Open circles stand for A$_{1g}$ mode. Close
circle stand for Raman shift of E$_{1g}$ mode obtained by
extrapolating the temperature dependent Raman shift of E$_{1g}$ mode
below Na ordering transition temperature to room temperature for
x=0.56 sample. Open triangle and diamond data come from Ref[21] and
Ref[22], respectively. Dark yellow line stand for O1 ordering region
phase defined by Huang et al.. Dotted dark yellow line stands for Na
reordering region defined in this paper. The black and red dash line
are guided for eyes.} \label{fig3}
\end{figure}

In order to further study the possible Na rearrangement transition,
we chose x=0.56 sample and measured temperature dependent Raman
shift (80 K $\sim$ 310 K). In Fig.4, temperature dependent Raman
spectrum for x = 0.56 shows different temperature dependence for
A$_{1g}$ and E$_{1g}$ mode. The temperature dependent Raman shift of
A$_{1g}$ and E$_{1g}$ mode is shown in Fig.4(a). As shown in
Fig.4(b) and (c), A$_{1g}$ mode linearly changes with decreasing
temperature. Although a step-like change happens around $T_s \sim
240$ K for the E$_{1g}$ mode, Raman shift of the E$_{1g}$ mode
follows a T-linear dependence below and above $T_s$. According to
the temperature dependence of Raman shift below $T_s$, a Raman shift
could be obtained by extrapolating to room temperature. As shown in
Fig.3, the Raman shift of E$_{1g}$ mode obtained by extrapolating to
room temperature falls to the abnormal shift region around x=0.5 in
Na ordering O1 structure. Since the sample is placed in a airtight
box for low temperature measurements, the signal-to-noise ratios of
spectra are much lower than those in Fig. 2. We can not see the weak
Raman peak around 425 cm$^{-1}$ below 240 K, which exist in x=0.5
and 0.52 sample. Some previous data are also included in Fig.3, it
further indicates a Na ordering transition from O1 to H1 structure
at $T_s \sim 240$ K with increasing temperature for the sample with
x=0.56.  Detailed information about this kind of Na ordering
transition needs to confirm  by further neutron or electron
scattering experiment.

\begin{figure}[h]
%\captionstyle{flushleft} \onelinecaptionsfalse
\centering
\includegraphics[width=0.4\textwidth]{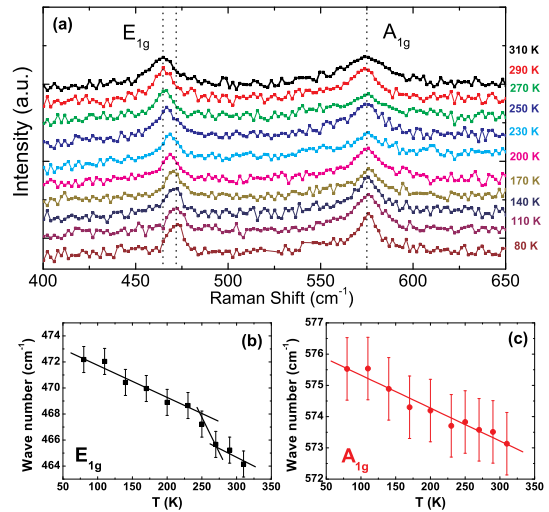}
\caption{(color online).(a): Temperature dependent Raman spectra
of Na$_x$CoO$_2$ single crystal with x=0.56; Temperature dependent
Raman Shift of (b): E$_{1g}$ mode and (c): A$_{1g}$ mode for
Na$_x$CoO$_2$ single crystal with x=0.56.} \label{fig4}
\end{figure}

\begin{figure}[h]
%\captionstyle{flushleft} \onelinecaptionsfalse
\centering
\includegraphics[width=0.4\textwidth]{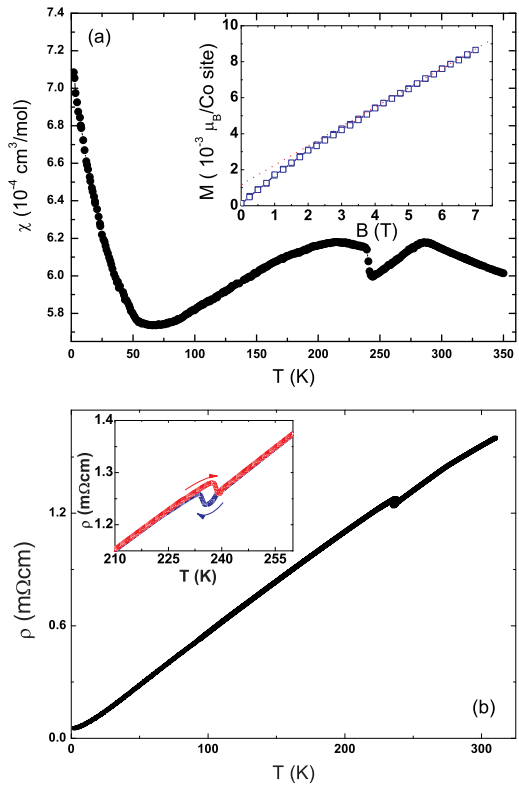}
\caption{(color online). Temperature dependent (a): susceptibility
with B = 7 T and B $\parallel$ Co-O plane and (b): in-plane
resistivity of Na$_x$CoO$_2$ single crystal with x=0.56. The inset
in (a): magnetization as a function of magnetic field up to 7 T at
4 K. The inset in (b): temperature dependent resistivity around
structural transition temperature with cooling and heating
measurement.} \label{fig5}
\end{figure}

Temperature dependent resistivity and susceptibility for x = 0.56
are shown in Fig.5 in whole temperature region (2 K $\leq$ T $\leq$
300 K). In Fig.5(a), the susceptibility under H = 7 T parallel to
Co-O plane shows an anomaly around 240 K, being consistent with Na
reordering transition observed in Raman study. Above 280 K, the
temperature dependent susceptibility shows a Curie-Weiss behavior.
The data were fitted with the Curie-Weiss law:
$\chi$=$\chi$$_0$+$\frac{C}{T+\theta}$. The fitting parameter
$\theta$ is 124 K and number of S=1/2 local spin is 0.14/Co site.
This result is in agreement with the report for x=0.7
sample.\cite{Foo} Below 240 K, the susceptibility shows a Pauli
paramagnetism-like behavior. No antiferromagnetic ordering is
observed down to 4 K, unlike x=0.5 sample\cite{Foo}. A Curie tail
exists below 50 K. The Curie tail is fitted with Curie-Weiss law,
and the value of $\theta$ is 46 K and the number of S=1/2 local spin
is 0.03/Co site. It indicates that the efficient number of S=1/2
local spin and antiferromagnetic correlation are reduced due to the
Na reordering transition. A crossover from Curie-Weiss to Pauli
paramagnetism is accompanied with Na reordering transition from H1
to O1 structure. A linear behavior is observed in field dependent
susceptibility at 4 K and only a tiny nonlinear behavior is observed
below 2 T, as shown in inset of Fig.5(a). This also indicates that
no ferromagnetic ordering happens, the ferromagnetic ordering shows
up in x=0.55 and 0.52 sample.\cite{Wang, Wu} In Fig.5(b), the
temperature dependent resistivity shows a loop behavior around 240 K
and the metal behavior is observed in the entire temperature-region.
Below 30 K, the temperature dependent resistivity obeys T$^{1.5}$
law. Between 30 K and 200 K, a well defined T-linear behavior is
found. These characteristics are similar to that in Pauli
paramagnetic metal for x$<$0.5, being consistent with above magnetic
properties. Isothermal magnetoresistance and temperature dependent
magnetoresistance under 7 T are shown in Fig.6. A positive and
monotonic magnetoresistance is observed as shown in Fig.6(a) and
(b). Below 30 K, a prominent positive magnetoresistance is observed
under 7 T with magnetic field both parallel and perpendicular to
Co-O plane, which is matched with the T$^{1.5}$ behavior in
resistivity. The maximum of magnetoresistance reaches 24$\%$ and 11
$\%$ with H perpendicular and parallel to Co-O plane, respectively.
These results indicate that strong magnetic fluctuation occurs below
30 K and a novel paramagnetic state maybe exist at low temperature.

\begin{figure}[h]
%\captionstyle{flushleft} \onelinecaptionsfalse
\centering
\includegraphics[width=0.4\textwidth]{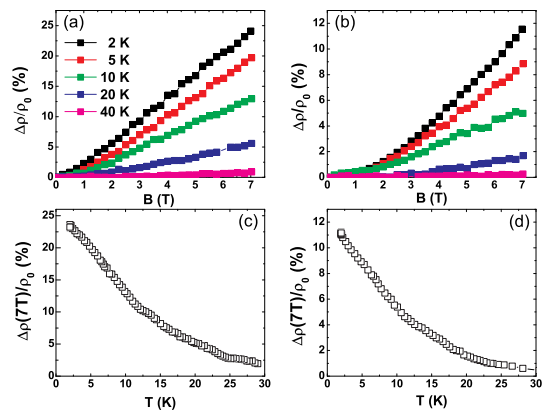}
\caption{(color online). Magnetic field dependent in-plane
magnetoresistance with external field (a) perpendicular to Co-O
plane and (b) parallel to Co-O plane. (c) and (d):
magnetoresistance at 7 Tesla as a function of temperature with
external field (c) perpendicular to Co-O plane and (d) parallel to
Co-O plane.} \label{fig6}
\end{figure}

Recently, more and more experiments indicated that Na ordering is
the key to understanding novel properties in Na$_x$CoO$_2$
system.\cite{Roger, Chou, Zhou, Marianetti, Choy} A plenty of Na
ordering pattern and rearrangement transition between these
different ordering patterns are found.\cite{Morris, Chou} But direct
evidence to proved the relationship between Na ordering pattern and
physical properties of Co-O plane is still limited.\cite{Schulze}
The above results give strong evidence that Na reordering transition
from H1 to O1 structure can induce a crossover from Curie-Weiss to
Pauli paramagnetic metal. Above Na reordering transition, a
Curie-Weiss behavior similar to that observed in x=0.7 sample is
confirmed by the temperature dependent susceptibility. Below Na
reordering transition, temperature dependent resistivity and
susceptibility are perfectly consistent with Pauli paramagnetic
metal for x$<$0.5 sample. These results indicate a direct link
between Na reordering transition and crossover from Curie-Weiss to
Pauli paramagnetic metal. There exists contradiction about the
abnormal "Curie-Weiss" metal for a long time. Recently, Shubnikov de
Haas effect has been studied in x=0.84 and 0.71
samples.\cite{Balicas} They found that the Na ordering can
reconstruct the Fermi surface (FS) pockets and leads to the
formation of local spin. In this picture, the crossover observed
here can be attributed to another kind of reconstruction of FS
pockets induced by Na reordering transition. Local spins are reduced
and Pauli paramagnetism is enhanced by this kind of reconstruction
of FS pockets. These results also shed light to the mechanism of
charge ordering insulator for x=0.5 sample. The charge ordering
insulator evolves from Pauli paramagnetic metal. Spin density wave
(SDW) or charge density wave(CDW) mechanism is favor for this
situation and early NMR result supports this picture.\cite{Bobroff}
But many recent experiments show that density wave mechanism can not
respond to insulator state for x=0.5 sample, and Na ordering effect
is proposed to understand it.\cite{Argyriou, Ning, Garbarino} The
present data can be understood by density wave mechanism, but if
there exists any other Na ordering pattern different from x=0.56 at
low temperature same as that in x=0.5, the density wave picture will
be challenged.

\section{Conclusion}
Raman spectroscopy of cleaved Na$_x$CoO$_2$ single crystals with
0.37 $\leq$ x $\leq$ 0.80 was systematically studied. The Raman
shift of A$_{1g}$ mode is found to be linearly dependent on Na
content. While the Raman shift of E$_{1g}$ mode has an abnormal
shift to high frequency around x = 0.5. The abnormal shift is
ascribed to Na$^+$ ordering in O1 structure.  Na reordering
transition from O1 to H1 structure around 240 K is proved by
temperature dependent Raman spectrum for x = 0.56 sample. Electronic
transport and susceptibility measurement for x=0.56 indicate that
the Na ordering in different structures leads to distinct physical
properties. Present results give a direct evidence for the
relationship between Na ordering and physical properties of Co-O
plane. Effect of Na ordering on physical properties is clearly
presented

This work is supported by the grant from the Nature Science
Foundation of China and by the Ministry of Science and Technology of
China (973 project No: 2006CB601001 and 2006CB922005), the Knowledge
Innovation Project of Chinese Academy of Sciences, and the
Innovation Foundation of USTC for the Postgraduate (KD2007077).


\begin{references}

\bibitem{Takada}
K. Takada et al., Nature (London) {\bf 422}, 53(2003).

\bibitem{Ong}
N. P. Ong and R. J. Cava, Science {\bf 305}, 52(2004).

\bibitem{Foo}
M. L. Foo et al., Phys. Rev. Lett. {\bf 92}, 247001(2004).

\bibitem{Zandbergen}
H. W. Zandbergen, M. L. Foo, Q. Xu, V. Kumar and R. J. Cava, Phys.
Rev. B {\bf 70}, 024101(2004).

\bibitem{Huang}
Q. Huang, M. L. Foo, J. W. Lynn, H. W. Zandbergen, G. Lawes, Yayu
Wang, B. H. Toby, A. P. Ramirez, N. P. Ong, and R. J. Cava (2004),
cond-mat/0402255.

\bibitem{Huang2}
Q. Huang, M. L. Foo, R. A. Pascal Jr., J. W. Lynn, B. H. Toby, Tao
He, H. W. Zandbergen, and R. J. Cava (2004), cond-mat/0406570.

\bibitem{Huang3}
Q. Huang, B. Khaykovich, F.C. Chou, J.H. Cho, J. W. Lynn, Y. S.
Lee, Phys. Rev. B {\bf 70}, 134115(2004).

\bibitem{Roger}
M. Roger, D. J. P. Morris, D. A. Tennant, M. J. Gutmann, J. P.
Goff, J.-U. Hoffmann, R. Feyerherm, E. Dudzik, D. Prabhakaran, A.
T. Boothroyd, N. Shannon, B. Lake and P. P. Deen, Nature (London)
{\bf 445} 631-634(2007).

\bibitem{Schulze}
T. F. Schulze, P. S. H\"{a}fliger, Ch. Niedermayer, K.
Mattenberger, S. Bubenhofer, and B. Batlogg, Phys. Rev. Lett. {\bf
100} 026407(2008).

\bibitem{Morris}
D. J. P. Morris et al., arXiv:0803.1312v2

\bibitem{Chou}
F. C. Chou et al., arXiv:0709.0085v1

\bibitem{Balicas}
L. Balicas, Y. J. Jo, G. J. Shu, F. C. Chou, P. A. Lee, Phys. Rev.
Lett. {\bf 100}, 126405(2008).

\bibitem{Pai}
Woei Wu Pai, S. H. Huang, Ying S. Meng, Y. C. Chao, C. H. Lin, H.
L. Liu, F. C. Chou, arXiv:0805.0475v1

\bibitem{Zhou}
Sen Zhou and Ziqiang Wang, Phys. Rev. Lett. {\bf 98},
226402(2007).

\bibitem{Marianetti}
C. A. Marianetti and G. Kotliar, Phys. Rev. Lett. {\bf 98},
176405(2007).

\bibitem{Choy}
Ting-Pong Choy, Dimitrios Galanakis, and Philip Phillips, Phys.
Rev. B {\bf 75}, 073103(2007).


\bibitem{Wu}
T. Wu, D. F. Fang, G. Y. Wang, L. Zhao, G. Wu, X. G. Luo, C. H.
Wang and X. H. Chen, Phys. Rev. B {\bf 76}, 024403(2007).

\bibitem{Shi}
Y. G. Shi, Y. L. Liu, H. X. Yang, C. J. Nie, R. Jin, and J. Q. Li,
Phys. Rev. B {\bf 70}, 052502(2004).

\bibitem{Iliev}
M. N. Iliev, A. P. Litvinchuk, R. L. Meng, Y. Y. Sun, J.
Cmaidalka, and C. W. Chu, Physica C {\bf 402}, 239(2004).

\bibitem{Lemmens}
P. Lemmens, V. Gnezdilov, N. N. Kovaleva, K. Y. Choi, H. Sakurai,
E. Takayama-Muromachi, K. Takada, T. Sasaki, F. C. Chou, D. P.
Chen, C. T. Lin, and B. Keimer, J. Phys.: Condens. Matter {\bf
16}, S857(2004).

\bibitem{Qu}
J. F. Qu, W. Wang, Y. Chen, G. Li, and X. G. Li, Phys. Rev. B {\bf
73}, 092518(2006).

\bibitem{Zhang}
Qingming Zhang, Ming An, Shikui Yuan, Yong Wu, Dong Wu, Jianlin
Luo, Nanlin Wang, Wei Bao, and Yening Wang, Phys. Rev. B {\bf 77},
045110(2008).

\bibitem{Lemmens2}
P. Lemmens, K. Y. Choi, V. Gnezdilov, E. Ya. Sherman, D. P. Chen,
C. T. Lin, F. C. Chou, and B. Keimer, Phys. Rev. Lett. {\bf 96},
167204(2006).

\bibitem{Donkov}
A. Donkov, M. M. Korshunov, I. Eremin, P. Lemmens, V. Gnezdilov,
F. C. Chou, and C. T. Lin, Phys. Rev. B {\bf 77}, 100504(R)(2008).


\bibitem{Wang}
C. H. Wang, X. H. Chen, T. Wu, X. G. Luo, G. Y. Wang, and  J. L.
Luo, Phys. Rev. Lett. {\bf 96}, 216401(2006).

\bibitem{Bobroff}
J. Bobroff, G. Lang, H. Alloul, N. Blanchard, and G. Collin, Phys.
Rev. Lett. {\bf 96}, 107201(2006).

\bibitem{Argyriou}
D. N. Argyriou, O. Prokhnenko, K. Kiefer, and C. J. Milne,
arXiv:0709.1038(2007).

\bibitem{Ning}
F. L. Ning, S. M. Golin, K. Ahilan, T. Imai, G.J. Shu and F. C.
Chou, arXiv:0711.4023(2007).

\bibitem{Garbarino}
G. Garbarino, M. Monteverde, M. Nunez Regueiro, C. Acha, M. L.
Foo, R. J. Cava, arXiv:0710.4341v1





\newpage

\noindent

\end{references}
\end{document}